\documentclass{article}
\usepackage{spconf,amsmath,graphicx}
\usepackage{textcmds}
\usepackage{booktabs}
\usepackage{multirow}
\usepackage{bm}
\usepackage{url}
\usepackage[backref=page]{hyperref}
\usepackage{cite}
\usepackage{subcaption}
\usepackage{enumitem}

\def\x{{\mathbf x}}

\title{Can we use Common Voice to train a Multi-Speaker TTS system?}
\name{Sewade Ogun, Vincent Colotte, Emmanuel Vincent}
\address{Université de Lorraine, CNRS, Inria, LORIA, F-54000 Nancy, France \\
sewade.ogun@inria.fr}
\copyrightnotice{978-1-6654-7189-3/22/\$31.00~\copyright~2023 IEEE\hfill}

\begin{document}

%\ninept
%
\maketitle
\begin{abstract}
Training of multi-speaker text-to-speech (TTS) systems relies on curated datasets based on high-quality recordings or audiobooks. Such datasets often lack speaker diversity and are expensive to collect. As an alternative, recent studies have leveraged the availability of large, crowdsourced automatic speech recognition (ASR) datasets. A major problem with such datasets is the presence of noisy and/or distorted samples, which degrade TTS quality. In this paper, we propose to automatically select high-quality training samples using a non-intrusive mean opinion score (MOS) estimator, WV-MOS. We show the viability of this approach for training a multi-speaker GlowTTS model on the Common Voice English dataset. Our approach improves the overall quality of generated utterances by 1.26 MOS point with respect to training on all the samples and by 0.35 MOS point with respect to training on the LibriTTS dataset. This opens the door to automatic TTS dataset curation for a wider range of languages.

\end{abstract}
\section{Introduction}
\label{sec:intro}

Research on text-to-speech (TTS) is increasingly focusing on multi-speaker TTS as it is more challenging and often requires explicit modelling of speaker characteristics. This interest has helped improve the performance of multi-speaker TTS in terms of prosody \cite{sun2020fully}, expressiveness \cite{valle2020mellotron}, new speaker generation \cite{stanton2022speaker}, zero-shot training  \cite{casanova2021sc}, and synthetic data generation for downstream tasks like automatic speech recognition (ASR) \cite{hu2022synt++}, among others. Depending on the application, different characteristics of speech need to be modeled. For example, for synthetic ASR training data generation, it is necessary for the model to have seen diverse speakers and accents. 

Datasets currently used for TTS system training fall into two categories, namely studio-quality TTS datasets such as VCTK \cite{yamagishi2019vctk}, and TTS datasets curated from audiobooks such as LibriTTS \cite{zen2019libritts}. However, the VCTK dataset includes only 110 speakers, and LibriTTS has a concentration of US English accents, which are not representative of the entire spectrum of speakers and accents. On top of that, the collection of datasets such as VCTK may be too expensive for some languages.

Large, crowdsourced ASR datasets are good candidates for driving TTS research in future directions, as they inherently exhibit the larger speaker variability (in terms of accent, speaking style, speaking rate, etc.) required for TTS systems to model diverse speakers. However, problems such as noise, low bandwidth, mispronunciation, variation in recording conditions, etc., hinder their usability for TTS training.

In this paper, we focus on automatically selecting high-quality training samples from a crowdsourced dataset, using Common Voice English \cite{ardila2019common} as an example.
In this context, quality cannot be estimated via subjective listening tests, that are intractable with 1.4~M utterances, or objective metrics like PESQ \cite{rix2001perceptual} that require a reference signal. 
% These measures have also been shown not to always correlate with subjective quality \cite{loizou2011speech, recommendation2001perceptual}.
Instead, we leverage the increasing accuracy of deep learning based, non-intrusive quality estimators. Specifically, we use a self-supervised model fine-tuned for mean opinion score (MOS) estimation, WV-MOS \cite{andreev2022hifi++}, and select the speakers whose average WV-MOS score across all utterances is above a threshold. %As this method is fully automatic, it could be applied to other crowdsourced speech datasets.
We evaluate the intelligibility, audio quality and speaker similarity of the utterances generated by a multi-speaker GlowTTS model trained on the resulting dataset, and also briefly explore the other factors not captured by WV-MOS.

% Finally, we show using subjective and objective measures that, by training a multi-speaker GlowTTS model on the selected corpus, the utterances generated by this model can achieve similar quality and intelligibility to the utterances generated by a TTS model trained on relatively-clean audiobook corpus (LibriTTS).

%The remaining part of the paper is organised as follows. 
Section~\ref{sec:related_work} describes related works on TTS dataset creation, TTS training on noisy speech, and MOS estimation. Section~\ref{sec:dataset} describes the Common Voice dataset, its properties and limitations. Sections~\ref{sec:typestyle} and \ref{sec:exp_setup} describe our method and the experiments performed to validate it. We conclude in Section~\ref{sec:conc}.

\section{Related work}
\label{sec:related_work}
Several multi-speaker datasets have been collected in recent years for TTS applications \cite{yamagishi2019vctk, zen2019libritts}. To create these corpora, researchers either record utterances in semi-anechoic chambers for good signal quality, or utilise various methods to select utterances from audiobooks, as this is less cumbersome. For example, the LibriTTS dataset was derived from the popular LibriSpeech ASR dataset \cite{panayotov2015librispeech} by trimming silences and filtering out utterances with low estimated signal-to-noise ratio (SNR). Although this filtering step is not perfect and allows a few noisy samples to remain uncaught, the resulting dataset is believed to be good enough for TTS since the original LibriSpeech is higher-quality than Common Voice on average.

A few works have used other metrics such as the word error rate or the Mel cepstral distortion to automatically select good training utterances from noisy speech datasets \cite{baljekar2016utterance, kuo2018data, cooper2017utterance}, however they have only been demonstrated on small datasets so far. Research on multi-speaker TTS training using noisy speech has also focused on directly modelling the noise in order to factor it out during inference \cite{zhang2021denoispeech, hsu2019disentangling}, and on encoding all the environmental characteristics of speech for novel speech generation in different conditions \cite{chang2021style}. 

Recently, several methods have been proposed to automatically measure the quality of speech utterances \cite{lo2019mosnet, patton2016automos} but they do not always generalise well outside of the training corpus. Self-supervised pretrained models followed by a shallow MOS regression head result in higher correlation with human evaluators' scores \cite{cooper2022generalization} than previous architectures. They also generalise better to unseen speakers and utterances, and can be used to evaluate the performance of speech processing systems for a variety of tasks \cite{andreev2022hifi++}.

\section{The Common Voice Dataset}
\label{sec:dataset}
Common Voice \cite{ardila2019common} is a crowdsourced, Creative Commons Zero licensed, read speech dataset currently available in over 93 languages. It contains recordings from volunteers who read a text transcript sourced from public domain text. Each utterance is up-voted or down-voted by volunteers according to a list of criteria.\footnote{\url{https://commonvoice.mozilla.org/en/criteria}} These criteria are not very restrictive, e.g., various kinds of background noises are allowed. Utterances with more than two up-votes are marked as validated. The validated utterances are then split into train, development and test sets, with non-overlapping speakers and sentences.

\subsection{Analysing Common Voice Dataset Quality for TTS}

Although the validated set has been widely exploited for ASR \cite{riviere2020unsupervised}, we observe some undesirable properties for TTS:
%\begin{list}{\labelitemi}{\leftmargin=1em}
\begin{itemize}[noitemsep,nolistsep]
\item Noise: Speech quality may be degraded by electromagnetic noise or acoustic noise such as mouse clicks, low frequency noise, background speakers and background music, among others. Since the utterances are stored as mp3, quantization noise can also sometimes be heard. 
\item Low bandwidth: Due to recording choices or high compression, some audio files are low-pass filtered, with a cutoff frequency that varies from one file to another.
\item Mispronunciation: We observe mispronunciations of \qq{unfamiliar} words, variations in the pronunciation of certain other words, and some utterances in other languages (e.g., German utterances in the English corpus).
\item Unavailable speaker metadata:  Age, gender and accent information are not available for all speakers, while some TTS systems require this information as input.
\item Other factors include variable recording characteristics (microphone, room, recording device), speaking rate, and volume. These recording characteristics must be ignored by models, and the speaking rate and volume, while being inherent characteristics of the speaker, can enlarge the space of variables to be considered.
%\end{list}
\end{itemize}
These characteristics are generally not a hindrance for ASR training, and they can even be desirable for robustness. However this is not the case for TTS training \cite{baljekar2016utterance, zen2019libritts}.
% High dataset uniformity and quality are two dataset properties that have been shown to improve the quality of utterances generated by TTS systems. 
%As such, we automatically filter the dataset by only selecting speakers with high estimated MOS scores. We believe utterances from these speakers are of high quality and are devoid of noise and missing frequency bands. To ascertain this, we train different TTS models on the same dataset filtered at different MOS thresholds. We further evaluate the importance of other issues listed above for a large-scale ASR dataset. -> NOT THE RIGHT PLACE, MOVED BELOW

\subsection{Dataset Preparation}
In the following, we use the English subset of Common Voice (version 7.0).
% \footnote{dataset can be freely downloaded from \url{https://commonvoice.mozilla.org/en/datasets}}
We exclude the predefined development and test sets, and utterances longer than 16.7~s to allow large batch sizes. We consider all other utterances in the 2015~h validated set as candidate TTS training samples. The samples are preprocessed by resampling from 32 or 48~kHz to 16~kHz, and removing beginning and end silences using pydub\footnote{\url{https://github.com/jiaaro/pydub}} with a threshold of -50~dBFS. The range of speaker duration is also limited to between 20~min and 10~h by randomly selecting a 10~h subset of utterances for speakers with longer duration and discarding speakers with less than 20~min total duration.

Furthermore, the training utterances are denoised using the pretrained DPTNet model of Asteroid \cite{pariente2020asteroid}. We run separate experiments for the original and denoised utterances to evaluate the impact of denoising on the resulting TTS model.

\section{Methodology}
\label{sec:typestyle}

We filter the dataset by only selecting speakers with high automatically estimated MOS scores. We believe that utterances from these speakers are of high quality and  devoid of noise and missing frequency bands. To ascertain this, we train different TTS models on the same dataset filtered at different estimated MOS thresholds. %We further evaluate the importance of other issues listed above for a large-scale ASR dataset. -> REMOVED BECAUSE NOT MENTIONED IN THIS SECTION

\subsection{MOS Estimation}
MOS estimation is performed using WV-MOS \cite{andreev2022hifi++}, a pretrained MOS estimation model.\footnote{\url{https://github.com/AndreevP/WV-MOS}} The model combines a pretrained wav2vec2.0 feature extractor and a 2-layer multi-layer perceptron (MLP) head, which are jointly fine-tuned on the subjective evaluation scores of the Voice Conversion Challenge 2018 using a mean squared error loss. It was shown to correlate well with human quality judgment regarding noise and low bandwidth \cite[App.~C]{andreev2022hifi++}. 
% The predicted MOS scores also had a high correlation with crowd-sourced MOSes, and had better system-level correlations with results than other objective metrics used in their evaluation. 

Every speaker is assigned a single, speaker-level WV-MOS score by averaging the estimated utterance-level scores. We assume that recording and environmental conditions for each speaker remain relatively constant. %, as shown by the small standard deviation of utterance-level WV-MOS scores. 
% This is in line with the finding in \cite{lee2018comparison} that speaker-level selection works better than utterance-level selection for TTS (albeit using a different selection metric).

We select all utterances from those speakers whose speaker-level WV-MOS score is above a threshold of 4.0, 3.8, 3.5, 3.0, or 2.0, and compare the resulting TTS systems with a baseline trained on all available data.\footnote{In the case of denoised training samples, the WV-MOS score is computed before denoising so that the list of selected speakers is not affected.} Table~\ref{tab:mos_table} shows the training data duration and number of speakers corresponding to each WV-MOS threshold. The thresholds were selected so as to balance data size and number of speakers.

\begin{table}[h!]
\caption{Training data duration and number of speakers for various selected WV-MOS thresholds.}

\label{tab:mos_table}
\centering
\begin{tabular}{ccc}
\hline
 WV-MOS threshold & Duration (h) & Number of speakers \\
 \hline
 Baseline & 636.27 & 633  \\
 $\text{WV-MOS} \geq 2.0$ & 620.14 & 623  \\
 $\text{WV-MOS} \geq 3.0$ & 532.14 & 537  \\
 $\text{WV-MOS} \geq 3.5$ & 310.40 & 337 \\
 $\text{WV-MOS} \geq 3.8$ & 187.40 & 183 \\
 $\text{WV-MOS} \geq 4.0$ & 86.05 & 88
\end{tabular}
\end{table}

% \vspace{-10}
\subsection{TTS Model}
We evaluate the dataset quality for TTS training at each WV-MOS threshold by training a multi-speaker GlowTTS model \cite{kim2020glow}, conditioned on an external speaker embedding, similar to \cite{casanova2021sc}. This model uses a Transformer encoder and a flow-based decoder, along with a phoneme duration prediction network. The model choice was influenced by its quality and its relatively short training time compared to other TTS models. Each utterance's speaker embedding and the corresponding sentence converted into phonemes are used as inputs to the model during training. The output is a Mel-spectrogram.

For each utterance, we pre-compute a speaker embedding from a speaker verification model\footnote{\url{https://github.com/resemble-ai/Resemblyzer}} trained on Voxceleb. The embeddings are l2-normalised, 256-dimensional vectors.

Lastly, a 16~kHz HiFi-GAN $V1$ vocoder \cite{kong2020hifi} was trained on the LibriTTS dataset to convert the generated Mel-spectro\-grams into audio signals. The vocoder was fixed and used to evaluate all the TTS systems considered, as the relative trends were true for vocoders trained on different datasets in our preliminary experiments. As such, the trained vocoder was not finetuned on Mel-spectrograms generated by GlowTTS so as to objectively evaluate the generated Mel-spectrograms. All experiments were carried out using the NeMo toolkit \cite{kuchaiev2019nemo}.

\section{Experimental Evaluation}
\label{sec:exp_setup}

%This section presents the experimental setup and results.% for our method. %First, we discuss the hyper-parameters and evaluation methods, then experimental results are presented.

\subsection{Training Hyper-Parameters}

All TTS models are trained using a global batch size of 128 or 256 on 4 GPUs (the noisier datasets only learn when the batch size is large). The model is optimised using the RAdam optimizer, a learning rate of $0.001$, and a cosine-annealing scheduler with linear warm up steps of 6,000. Each model is trained until the validation loss stops decreasing for more than 10 epochs. We select 504 utterances randomly from each dataset for validation. At inference, the generation is done with a noise scale of 0.667 and length scale of 1.0, which are the best multi-speaker inference parameters reported in \cite{kim2020glow}.

\subsection{Objective Evaluation}
To evaluate the TTS systems objectively, we generate utterances for speakers seen and speakers unseen at training time. 80 speakers are randomly selected from the smallest subset ($\text{WV-MOS} \geq 4.0$) of Common Voice\footnote{By design, all speakers in that subset are also included in the lower WV-MOS threshold training subsets.} and from the VCTK corpus to represent seen and unseen speakers, respectively. For each speaker, a single speaker embedding is extracted from a reference utterance of that speaker, that is either a randomly selected utterance with duration longer than 2~s for seen speakers or the fifth utterance (SpeakerID\_005) as in \cite{casanova2021sc} for unseen speakers. 
The embeddings are used to generate 25 utterances for each speaker, using text sentences from the VCTK corpus with more than 20 words. This results in a total of $2,000$ test utterances for both seen and unseen speakers.

To measure the audio quality, speaker similarity, and intelligibility of the generated utterances, we compute the average WV-MOS score \textbf{(WV-MOS)}, the cosine similarity between the speaker embeddings of the generated and the reference utterances \textbf{(cos-sim)}, and the character error rate \textbf{(CER)}, respectively. The pretrained QuartzNet$15\x5$ model from the NeMo toolkit was used to compute the CER.

\subsection{Subjective Evaluation}
We also evaluate the TTS systems subjectively in terms of overall quality \textbf{(MOS)} (whether audio sounds natural, non-robotic, and noiseless), speaker similarity between the generated and reference utterances \textbf{(S-MOS)}, and intelligibility.  In total, 24 volunteers participated in the evaluation.

For MOS and S-MOS, we selected 2 unseen male speakers (p245 and p254) and 2 unseen female speakers (p231 and p250) from the VCTK corpus, and generated 5 utterances per speaker for all TTS models. 
%Since this evaluation aims to compare different systems, we are confident that the number of speakers does not affect the comparison in terms of the quality of generated utterances and speaker similarity, and therefore enables us to keep the listening test duration sizeable.
Volunteers were asked to listen to the utterances carefully and give an MOS score for each utterance on a 1--5 scale, then an S-MOS score on a 1--5 scale with respect to a VCTK reference utterance. They were asked to ignore the speaking rate in their scoring, as the VCTK reference speaker often speaks faster than the generated utterances.

To measure intelligibility, we apply the minimal pair approach \cite{hodge2011minimal}. A minimal pair is a pair of words that vary by only one phoneme, e.g., sea/she.\footnote{We select minimal pairs from \url{https://www.englishclub.com/pronunciation/minimal-pairs.htm}}. This approach helps identify deficiencies in phoneme generation by the TTS systems, and it is widely used by phoneticians.
% and makes it possible to compare the various systems considered while allowing the evaluators to be non-native.  
In the evaluation, a carrier audio containing one of the pairs is presented to the evaluator and the evaluator has to select the word heard among three options: the correct word (e.g., sea), its minimal pair (e.g., she) or none of these. The proportion of correctly associated words is then computed as the \textbf{intelligibility score}. 25 minimal pairs were evaluated for each model considered.

\subsection{Results}
\subsubsection{Impact of the WV-MOS Threshold}
\begin{figure}[t!]
		\centering
	\begin{subfigure}{\linewidth}
        \includegraphics[width=8.5cm,trim=0 0 0 42,clip]{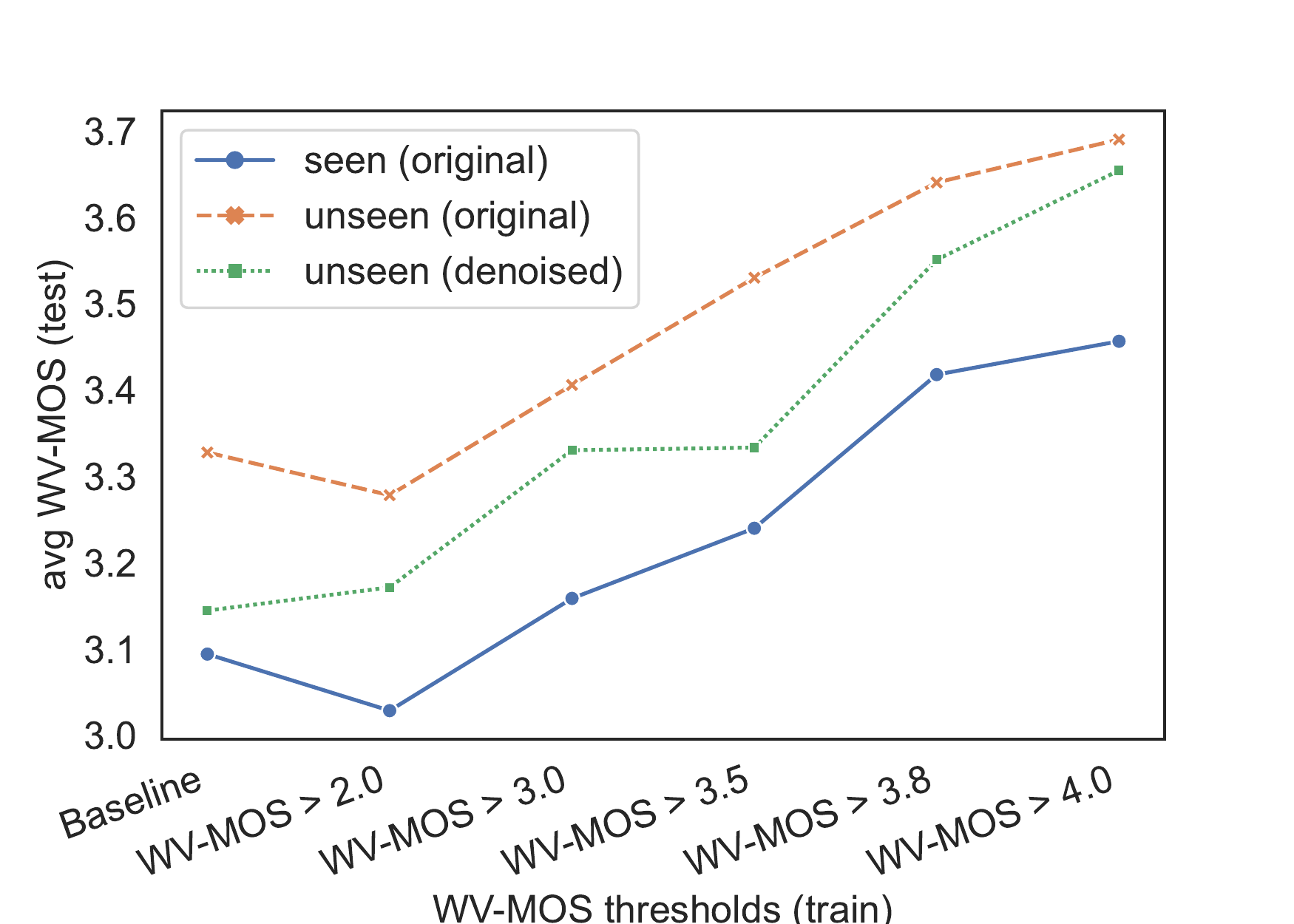}
        \vspace{-2pt}
        \caption{WV-MOS scores for seen and unseen speakers.}
        \label{fig:res1a}
    \end{subfigure}
    \begin{subfigure}{\linewidth}
	    \includegraphics[width=8.5cm,trim=0 15 0 20,clip]{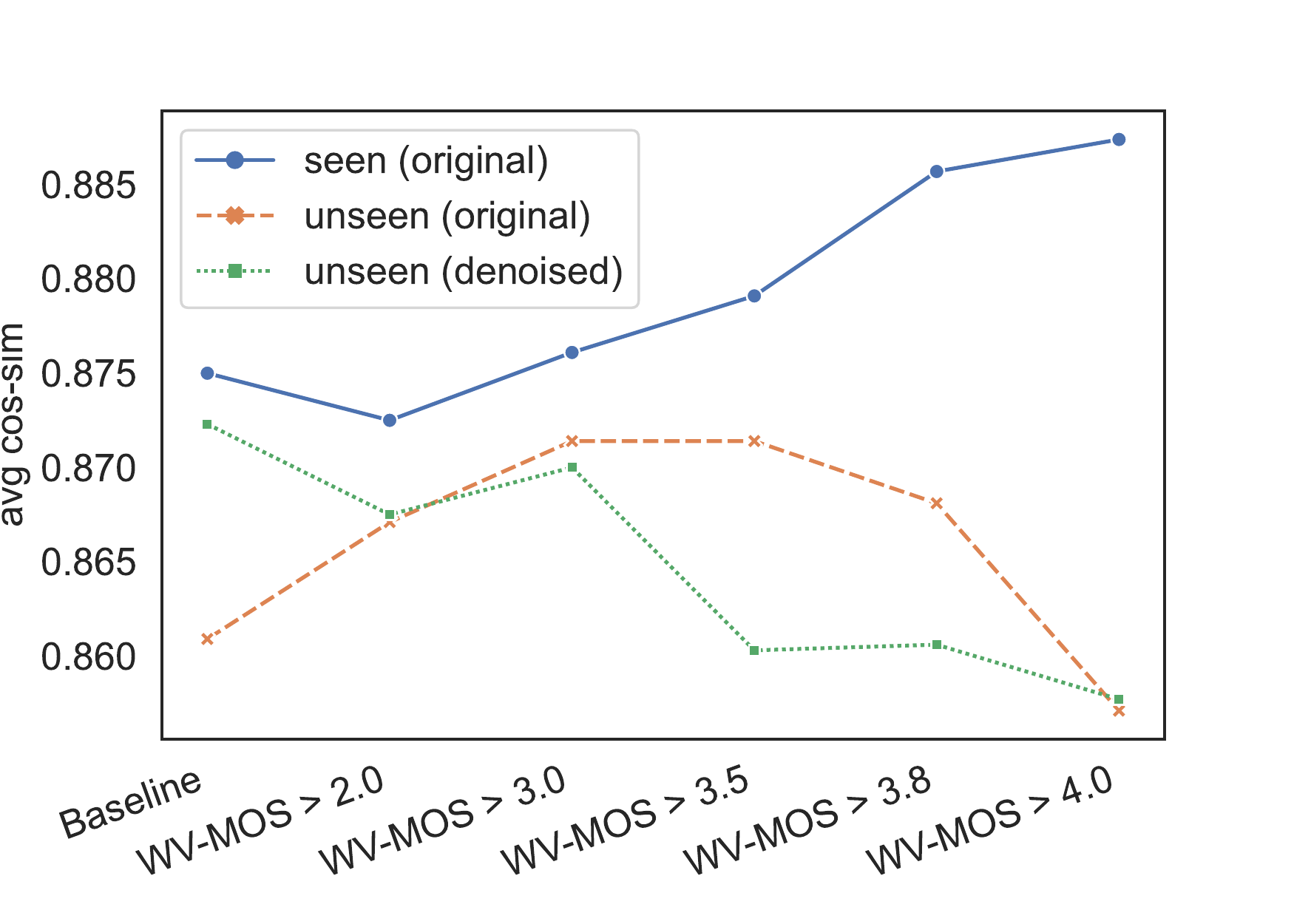}
        \vspace{-2pt}
        \caption{Cosine similarity between speaker embeddings.}
		\label{fig:res1b}
    \end{subfigure}
    \begin{subfigure}{\linewidth}
	    \includegraphics[width=8.5cm,trim=0 15 0 20,clip]{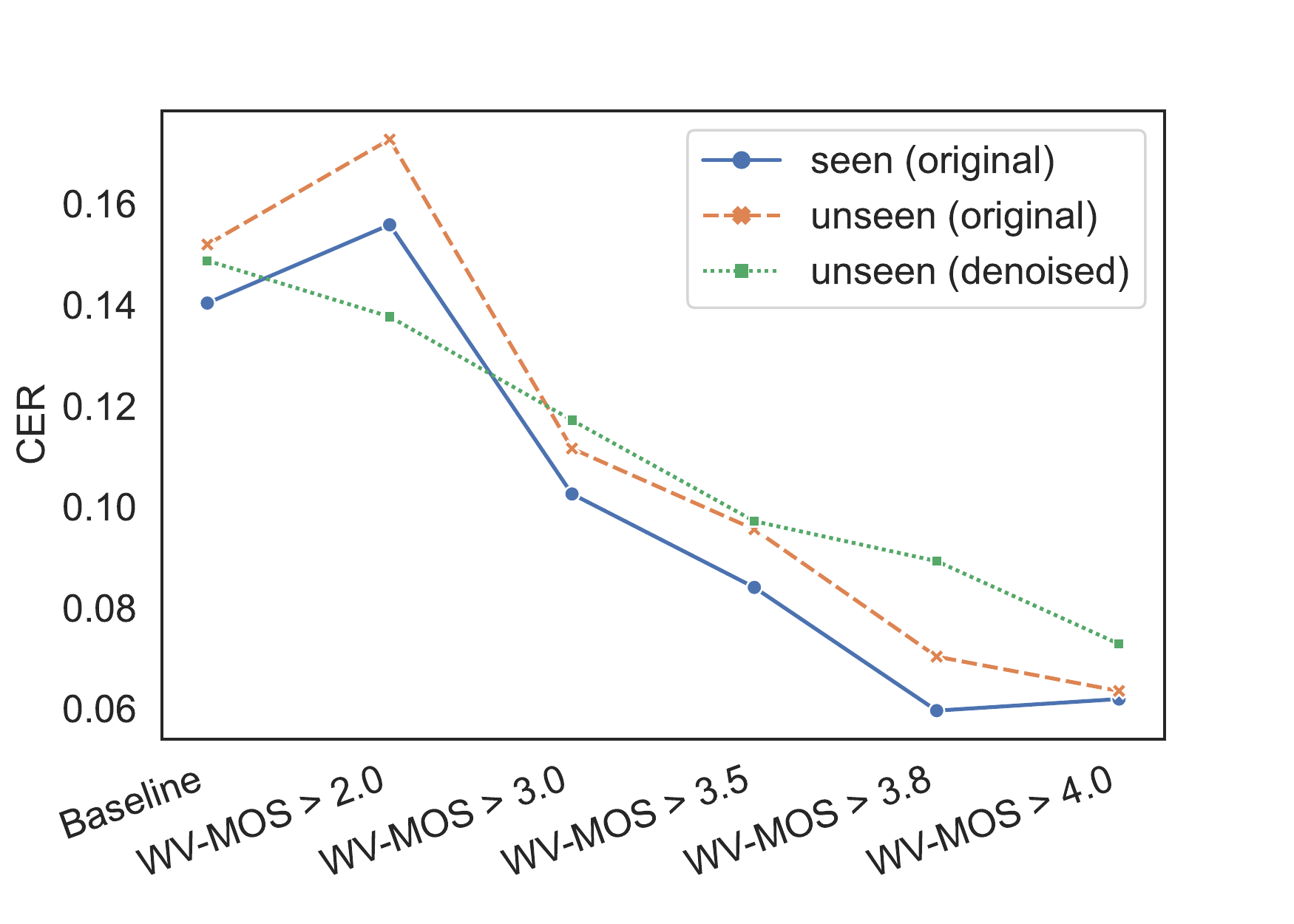}
        \vspace{-2pt}
        \caption{CER for seen and unseen speakers.}
		\label{fig:res1c}
    \end{subfigure}
	\caption{Objective quality (WV-MOS), speaker similarity (cos-sim) and intelligibility (CER) of utterances generated for 80 seen and 80 unseen speakers by TTS models trained on original or denoised samples above a given WV-MOS threshold.}
	\label{fig:res1}
\end{figure}

Figure \ref{fig:res1} shows the objective evaluation plots. 
The WV-MOS scores in Fig.\ \ref{fig:res1a} follow an increasing trend from the baseline to the $\text{WV-MOS} \geq 4.0$ dataset. Subjective MOS results in Table \ref{tab:mos_table_2} also corroborate this. This indicates that filtering based on WV-MOS scores improves the quality as expected.

In Fig.\ \ref{fig:res1b}, for models trained on original data, the cos-sim for seen speakers shows an increasing trend, while the cos-sim for unseen speakers is lowest at the two ends of the plot. We see a similar trend in Table \ref{tab:mos_table_2} for the S-MOS of unseen speakers. This indicates that more training speakers are key for modelling speaker variability.

In Fig.\ \ref{fig:res1c}, the CER decreases steadily from the noisy baseline dataset to the less noisy WV-MOS datasets for both seen and unseen speakers. We see a reduction of more than 50~\% in the CER from the baseline to the $\text{WV-MOS} \geq 4.0$ dataset. This indicates that intelligibility is mostly affected by the quality of the dataset, not the size.

The lowest quality datasets (Baseline vs.\ $\text{WV-MOS} \geq 2.0$) do not follow these trends due to the higher variance in utterance-level WV-MOS scores for lower-quality speakers. %in $[2.0, 3.0]$ speakers. Therefore, the $\text{WV-MOS} \geq 2.0$ dataset still include a high number of lower score utterances.

\begin{table}[h!]
\caption{Subjective / objective quality (MOS / WV-MOS) and speaker similarity (S-MOS / cos-sim) of utterances generated for 4 unseen speakers by TTS models trained on the baseline dataset and the $\text{WV-MOS}\geq3.0$ and $\text{WV-MOS}\geq4.0$ datasets. Bold numbers denote the best system in each row and the systems statistically equivalent to it.}

\label{tab:mos_table_2}
\centering
\begin{tabular}{|@{\hspace{3pt}}l|c@{\hspace{9pt}}c@{\hspace{9pt}}c@{\hspace{3pt}}|}
\cline{2-4}
\multicolumn{1}{l|}{}  & Baseline        & $\text{WV-MOS}\geq3.0$ & $\text{WV-MOS}\geq4.0$ \\ \hline
MOS     & $2.35$       & $3.12$            & $\mathbf{3.69}$            \\
WV-MOS   & $3.63$       & $3.59$            & $\mathbf{4.09}$            \\ \hline
S-MOS   & $2.69$       & $\mathbf{2.90}$            & $\mathbf{2.79}$            \\
cos-sim & $\mathbf{0.831}$ & $\mathbf{0.845}$      & $\mathbf{0.832}$     \\ \hline
\end{tabular}
\end{table}

\begin{table*}[ht!]
\centering
\caption{Subjective quality (MOS), speaker similarity (S-MOS) and intelligibility of utterances generated for 4 unseen speakers by TTS models trained on the baseline dataset, LibriTTS, and the $\text{WV-MOS}\geq4.0\text{-all}$ dataset. Corresponding objective scores (WV-MOS and cos-sim) are included. Bold numbers denote the
best system in each row and the systems statistically equivalent to it. Copy-synthesis on VCTK speech (VCTK-copy) provides an upper bound on the achievable speaker similarity.}
 \label{tab:mos_table_3}
\begin{tabular}{|ll|c|c|c||c|}
\cline{3-6}
\multicolumn{2}{l|}{}  & Baseline        & LibriTTS        & $\text{WV-MOS}\geq4.0\text{-all}$  & VCTK-copy        \\ \hline
\multirow{3}{*}{MOS}  & Male     & $2.44$       & $3.15$       & $\mathbf{3.70}$               & -               \\
                      & Female     & $2.26$       & $\mathbf{3.38}$       & $\mathbf{3.52}$             & -               \\
                      & Total & $2.35$       & $3.26$       & $\mathbf{3.61}$               & -               \\ \hline 
\multicolumn{2}{|l|}{WV-MOS}      & $3.63$       & $\mathbf{3.75}$       & $\mathbf{3.80}$               & -               \\ \hline\hline
\multirow{3}{*}{S-MOS} & Male     & $\mathbf{2.92}$       & $\mathbf{2.95}$       & $\mathbf{3.02}$               & $4.50$       \\
                      & Female     & $2.46$       & $2.53$       & $\mathbf{2.73}$               & $4.73$       \\
                      & Total & $2.69$       & $2.74$       & $\mathbf{2.88}$               & $4.61$       \\ \hline
\multicolumn{2}{|l|}{cos-sim}      & $0.831$ & $\mathbf{0.861}$ & $\mathbf{0.861}$         & $0.869$ \\ \hline\hline
\multicolumn{2}{|l|}{Intelligibility score}     & $0.72$                & $\mathbf{0.82}$            & $\mathbf{0.82}$                    & -               \\\hline %\hline 
%Duration in hours*             & -      & 636.27          & 492.68          & 230.75                  & -               \\ \hline
%Number of speakers              &  -     & 633             & 2484            & 4469                    & -               \\ \hline
\end{tabular}
\end{table*}

\subsubsection{Impact of Denoising Training Utterances}
As seen in Figs.\ \ref{fig:res1a} and \ref{fig:res1c}, the WV-MOS scores and CER for unseen speakers follow the same trend, whether the training data are denoised or not. 
%In this section, we will compare the various metrics for unseen test speakers. In Fig.\ \ref{fig:res1a}, for both original and denoised training samples, the test WV-MOS scores follow an increasing trend from the baseline to the $\text{WV-MOS} \geq 4.0$ dataset. We also see that the CER decreases from the baseline to the $\text{WV-MOS} \geq 4.0$ dataset in Fig.\ \ref{fig:res1c}.
Furthermore, denoising degrades the WV-MOS score and the CER, except for low quality datasets in the case of the CER.
In Fig.\ \ref{fig:res1b}, we see that denoising also degrades the cos-sim score, except for the baseline. This shows that some speaker information may be lost during denoising.
We conclude that denoising is not beneficial and automatically selecting high-quality samples is the best strategy.

\subsubsection{Common Voice vs.\ LibriTTS}
We compare our dataset curation method to a standard TTS dataset: LibriTTS. Since LibriTTS has more speakers (2,484) and more data (492.68~h after discarding utterances longer than 16.7~s) than the $\text{WV-MOS} \geq 4.0$ dataset, we select all speakers with WV-MOS above 4.0, without setting a 20~min lower bound on total speaker duration. We call the resulting 4,469-speaker, 230.75~h dataset $\text{WV-MOS} \geq 4.0\text{-all}$. Table \ref{tab:mos_table_3} evaluates the utterances generated by the TTS models trained on this dataset vs.\ LibriTTS for unseen speakers. 

Training on $\text{WV-MOS} \geq 4.0\text{-all}$ results in a similar intelligibility score to training on LibriTTS. 

The S-MOS score is highest when training on $\text{WV-MOS} \geq 4.0\text{-all}$, however we still see a large gap in S-MOS between this model and the VCTK-copy topline, which leaves room for further improvement in speaker modelling. 
% Also, the cos-sim is not statistically different between the two datasets >> Removed for space.
% The speaker similarity metrics cos-sim and S-MOS are also not statistically different across the two datasets. 
We notice that male S-MOS scores are higher than female S-MOS scores for both datasets, which indicates that male speakers are better modelled by models trained on either dataset.

Finally, while the WV-MOS scores for utterances generated by training on $\text{WV-MOS} \geq 4.0\text{-all}$ vs.\ training on LibriTTS are not statistically different, volunteers consistently gave higher MOS scores to the former, with an average improvement of 0.35 MOS point. Male speakers are assigned higher MOS scores, unlike LibriTTS where female voices are given higher MOS scores (in line with \cite{zen2019libritts}). %In terms of speaker similarity, training on the $\text{WV-MOS} \geq 4.0\text{-all}$ subset also improves the S-MOS for both male and female speakers, over LibriTTS dataset, highlighting the importance of having more speakers in the dataset. -> CAN'T SAY THIS UNLESS IT'S STATISTICALLY SIGNIFICANT

% Table \ref{tab:mos_table_3} shows the results comparing this dataset and LibriTTS across all metrics. Volunteers consistently rate utterances generated by "$\text{WV-MOS} \geq 4.0~\text{all}$" higher than those of LibriTTS or the baseline model. Also, male speakers are assigned higher scores by both Common Voice subsets, unlike LibriTTS where female voices are rated higher in terms of MOS (This is in line with results on LibriTTS \cite{zen2019libritts}). Training on the "$\text{WV-MOS} \geq 4.0~\text{all}$" subset also improves the S-MOS for both male and female speakers.
% % It is also observed that male S-MOS scores are higher than the female S-MOS for all datasets, except for VCTK test utterances with copy synthesis.
% LibriTTS and "$\text{WV-MOS} \geq 4.0~all$" have similar intelligibility scores, while the baseline is less intelligible to volunteers.

% \subsubsection{Other factors not captured by WV-MOS}
% We performed further experiments (not shown here) on the $\text{WV-MOS} \geq 3.5$ dataset to evaluate the effect of a) removing foreign languages b) filtering WV-MOS at the utterance level, instead of speaker-level c) removing utterances with low bandwidth d) removing long silences in the middle of utterances e) dropping utterances with very high CER. We did not observe any significant improvement in the WV-MOS scores and the CERs of test utterances. We attribute this to the fact that the WV-MOS score is robust and was able to consider many of these properties in its MOS estimation.

\subsubsection{Other factors not captured by WV-MOS}
We performed experiments on a medium quality dataset, $\text{WV-MOS} \geq 3.5$, to assess the impact of other factors. First, we removed sentences in a foreign language by using a language identification tool, LangID \footnote{\url{https://github.com/saffsd/langid.py}}, to filter out sentences with English language probability lower than 0.8. %About 6 hours of utterances were removed.
Second, using the same pretrained ASR model as above, we dropped utterances with a CER above 0.4 for better alignment of training text to utterances. %Note that utterances of the foreign language also had very high CER, therefore they were mostly excluded.
Third, we filtered utterances according to their WV-MOS at the utterance level, instead of the speaker level.
Finally, we removed pauses longer than 180~ms inside utterances using a voice activity detector.\footnote{\url{https://github.com/wiseman/py-webrtcvad}} Each of these experiments resulted in discarding less than 1.5 \% of the initial dataset. Informal listening tests did not show any improvement in quality and intelligibility. Although this would have required more formal tests to validate, we conclude that the other factors not captured by WV-MOS are not critical.

% Also, we did not observe any significant improvement in the WV-MOS scores and the CERs of test utterances, except when removing pauses. Informal listening tests did not validate this, however, as the utterances were less natural, we conclude that the other factors not captured by WV-MOS are not critical. >> # removed this as suggested

\section{Conclusion}
\label{sec:conc}

In this paper, we successfully improved the overall quality, speaker similarity, and intelligibility of utterances generated by a multi-speaker TTS model trained on the Common Voice English dataset. This was achieved by selecting high-quality training samples using a non-intrusive MOS estimator. Furthermore, we showed that denoising reduces the CER and increases the speaker similarity score (cos-sim) of generated utterances when the dataset is noisy, but degrades performance otherwise. The resulting automatically curated dataset shows promise for future TTS experiments, as it outperforms LibriTTS in terms of both subjective quality and speaker similarity. The applied approach is generic and could enable the creation of TTS training datasets for languages for which manual curation is not financially viable. In future work,
we will report the impact of vocoder training data quality on the absolute performance of the system.

\section{ACKNOWLEDGMENTS}
\label{sec:ack}

Experiments presented in this paper were carried out using the Grid'5000 testbed, supported by a scientific interest group hosted by Inria and including CNRS, RENATER and several Universities as well as other organizations (see \url{https://www.grid5000.fr}).

% % -------------------------------------------------------------------------
%\pagebreak
\bibliographystyle{IEEEbib}
\bibliography{strings,refs}

\end{document}